\documentclass[aps,twocolumn,a4paper]{revtex4}
\usepackage{graphicx,color}
\usepackage{amsmath,amssymb}
\usepackage{enumerate}
\usepackage{subfigure}
\usepackage{tabularx}
\usepackage{subfigure}  
\usepackage[all]{xy}
\usepackage{mathtools}
\newcommand{\be}{\begin{equation}}
\newcommand{\ee}{\end{equation}}
\newcommand{\ben}{\begin{eqnarray}}
\newcommand{\een}{\end{eqnarray}}
\newcommand{\bes}{\begin{subequations}}
\newcommand{\ees}{\end{subequations}}

\newcommand{\bb}{\bibitem}

\begin{document}
\title{{Configurational Entropy for Skyrmion-like Magnetic Structures}}
\author{D. Bazeia$^1$, D. C. Moreira$^2$ and E. I. B. Rodrigues$^3$}
\affiliation{$^1$Departamento de F\'\i sica, Universidade Federal da Para\'\i ba, 58051-970 Jo\~ao Pessoa, PB, Brazil\\
$^2$Unidade Acad\^emica de F\'\i sica, Universidade Federal de Campina Grande, 58109-970 Campina Grande, PB, Brazil\\
$^3$Universidade Federal Rural de Pernambuco, 54518-430 Cabo de Santo Agostinho, PE, Brazil}
\begin{abstract}
In this work we explore the relationship between two ideas recently introduced in the literature. The first one deals with a quantity related to the informational contents of solutions of spatially localized structures, and the second consists of obtaining analytical solutions to describe skyrmion-like structures in magnetic materials. In particular, we use the topological charge density to extract information on the configurational entropy of the magnetic structure. 
\end{abstract}
\date{\today}
\pacs{...}
\maketitle
\section{Introduction}

Skyrmions are topologically protected structures which appear in physics in several distinct environments. The idea arose in the context of hadron physics \cite{sky1,sky2}, where a meson model provided topological solitons, identified with baryons, whose topological charge was an integer, identified with the baryon number (for an extended discussion of this model, see \cite{zahed}). Skyrmions are also  being treated in the context of low-energy QCD as an effective theory both to explain the behavior of light baryons and to find solutions with topological charges larger than unity \cite{witten,sut3,sut4,sut2,sut1}. Another interesting feature on this is that one can extend the skyrmion Lagrangian simply by adding a sixth order term. This term together with a specific potential, isolated from the other Lagrangian contributions, leads to exact BPS solutions, since they saturate a Bogomol'nyi bound \cite{sanchez4,sanchez3,bps}. In this case the energy of the soliton solution is linearly related to the baryon charge and the binding energy is zero. This leads to the interpretation that the dominant contribution to the nuclear masses is given by the BPS solutions while contributions coming from other terms of the Lagrangian provides only small corrections \cite{sanchez2,sanchez}.

In condensed matter, skyrmions has acquired a prominent role after its theorization \cite{bog89,Bog3} and subsequent observation \cite{Binz,Yu,jiang} in experiments related with chiral magnetic systems, boosting the study of what nowadays is called {\it magnetic skyrmion}. In general lines, a magnetic skyrmion is a structure with rotational symmetry in which the magnetization at its center points upward (downward), and smoothly changes to the downward (upward) orientation as $r$ increase to higher values. Recently, such structures have been studied in several distinct environments, as confining structures in liquid crystals \cite{fukuda2011,Bog1,ackerman2014}, superconductors \cite{Garaud,Agterberg} and in other magnetic materials \cite{Bog2,seki,fert}. Skyrmions have also gained relevance for the coding and transmition of information \cite{liu,Reyren}.

There are several mechanisms that contribute to generate skyrmions in magnetic materials which usually act together in the formation of such structures. We can cite as examples the long-ranged magnetic dipolar interactions, the frustraded exchange interactions and the four-spin exchance interaction, which are introduced as ways of avoiding Derrick's argument by breaking the spatial inversion symmetry \cite{Yu,jiang}. Another important way to deal with magnetic skyrmions is via the Dzyaloshinskii-Moriya interaction \cite{dm1,dm2}, where we have a Lagrangian with highly nonlinear dynamic terms. This route brings serious difficulties in the search for solutions and, in fact, is it very hard to find analytical solutions. However, recently an alternative method has been introduced in the literature that provides analytical skyrmion-like solutions \cite{DDR,jmmm,jmmm2}. It links the framework used when one sets up isolated magnetic skyrmion with some tools known in scalar field theories. In this approach the scalar field, which also saturates the Bogomol'nyi bound, is derived from a model constructed to avoid Derrick's scaling argument \cite{8} and has a role in the orientation angle present in the magnetization of the skyrmion. There, the vorticity and helicity are adjusted in order to obtain skyrmions/vortex-like solutions, depending of the value of the topological charge. 

We have also to point out that by the prism of topological aspects, skyrmions and magnetic bubbles are equivalent. However, they differ qualitatively in several respects, as in their characteristic sizes (magnetic bubbles are much larger than skyrmions), the lifetime at room temperature (the lifetime of the magnetic bubbles is longer than the lifetime of the skyrmions) and the types of interactions that govern their behavior. While the main contribution to the existence of magnetic bubbles comes from the presence of an external magnetic field, the behavior of the skyrmions is mainly due to the dispute between spin exchange interactions and magnetic anisotropies \cite{nagaosa}.  Skyrmions may also appear as fundamental states in nanosystems with  spatial inversion asymmetry. This was theorized \cite{Bog3} and found \cite{heinze,beg} in models where the term Dzyaloshinskii-Moriya has a key role in reducing energy by favoring a single sense of orientation to the  skyrmion lattice. Finally, it has been shown that the construction of lattices of magnetic bubbles and skyrmions also depends on a set of topological constraints derived from the continuity of the magnetization vector and the boundary conditions imposed on it by the magnetic material composition \cite{bogatyrev2018}. Since the model that we approach in this work is constructed based on the topology of the solutions, it does not capture these differences. Thus, in this setup we can explore both skyrmions and magnetic bubbles or any other magnetic structure with the same topological features.

In order to find new ways to obtain results on the behavior of such magnetic structures, we will here implement the analysis from the point of view of a tool recently introduced in the context of Field Theory, which is called {\it configurational entropy} \cite{gleiser1}. Based on ideas related to Shannon entropy, this quantity is thought of as a way of obtaining a measure related to the informational content of the (possible) solution of a model, so it is presented as a way of estimating the {\it physical quality} of  spatially localized solutions, resulting from the non-linear differential equations used to describe the system. Calculated in the functional space, it associates the informational and dynamic contents of solutions with localized spatial energy. 

A particular property presented in this approach is that the higher the energy of a given ansatz to solve its system of equations, the higher is the configurational entropy. However, even in situations where different trial functions have degenerate energies, the configurational entropy acts as a tie-breaking factor in the choice of the quality of the solution. The procedure has been recently explored in some contexts such as Field Theory \cite{gleiser2,gleiser3,gleiser4,gleiser5,correa3,xxx1}, Statistical Mechanics \cite{gleiser6,gleiser8}, Gravitation \cite{gleiser7,correa2,correa4,roldao2,correa5} and the AdS/QCD correspondence \cite{roldao1,roldao22}. Thus, it is quite natural the interest in knowing how this quantity can be used in other areas as condensed matter. In particular, in the recent work \cite{gleiserX} the authors show the use of the configurational entropy as a interesting predictor of spontaneous decay rates for atoms.

Here, the tie-breaker is made in relation to the topological charge, since different constructions of skyrmions may have the same value for that quantity. The difference between them, in this case, lies in the distribution of the topological charge density over space, which may be more dispersed or more localized, and this certainly affects the way the information is distributed and quantified in those structures.

Guided by the above motivations, we organize the current work as follows. In Sec. II we present generalities about the construction of the models to be studied, the formalism and the obtaining of analytical solutions. In addition, we also present and discuss formally what is the configurational entropy and how to apply it to the skyrmion-like configurations. In Sec. III we apply these tools to discuss two distinct models, one already investigated in the recent literature, and the other new, both bringing novelties concerning the configuration entropy which is the main topic to be explored in this work. We end the investigation in Section IV with some comments and conclusions. 

\section{Generalities} 

Let us start reviewing some general features of the problem under consideration. Here we first deal with the topological features of the skyrmion-like structure and how it is described by the scalar field model, and then introduce the configurational entropy, linked to the topological charge density of such magnetic configuration.  

\subsection{Skyrmion-like structure} 

Magnetic skyrmions are planar topological objects characterized by the topological charge, also called skyrmion number, given by
\be\label{tc}
Q=\frac{1}{4\pi}\int d^2x\; {\bf M}\cdot\left(\partial_x{\bf M}\times\partial_y{\bf M}\right),
\ee
where {\bf M} is the unit vector associated with the magnetization of the structure. Such a charge may have integer or semi-integer value, normalized to assume $Q=\pm 1$ or $Q=\pm 1/2$. When (\ref{tc}) is integer we have skyrmion, and when it is half-integer, we have half-skyrmion or vortex-like solution.

We are interested in planar static solutions with rotational symmetry, so it is better to use cylindrical coordinates $(r,\theta,z)$ in such a description. As a consequence the magnetization has only radial dependence, i.e., ${\bf M}={\bf M}(r)$. Helical excitations requires the condition ${\bf M}\cdot{\hat r}=0$,  which allows us to write the magnetization as follows
\be\label{M}
{\bf M}(r)=(0,\cos\Theta(r) ,\sin\Theta(r)).
\ee 
Note that there is only one degree of freedom left, given by the angle $\Theta(r)$. We can, for convenience, rewrite this quantity as
 \cite{DDR,jmmm,jmmm2}
\be\label{T}
\Theta(r)=\frac{\pi}2\phi(r)+\delta,
\ee
where we separate the $\delta$-term, which is a phase associated with the helicity of the structure and,  in addition, the term $\pi/2$  is inserted for normalization of the skyrmion number (\ref{tc}). The radial dependence, in this way, is left to the scalar function  $\phi(r)$, which plays a key role in the model we study, acting as the quantity that in fact varies in space and connects the distinct ground states, delimiting the region covered by the defect. One can note that the values of the scalar function $\phi(r)$ at the boundaries and the phase $\delta$  must be self-tuned in  such a way that the topological charge \eqref{q} has the desired normalization. One can show that the cross product present in the integral \eqref{tc} can be rewritten in polar coordinates as
\begin{equation}
\partial_x \vec{M} \times \partial_y \vec{M} = -\frac{\pi}{2}\frac{\cos\Theta}{r}\frac{d\phi}{dr}\vec{M} 
\end{equation}
and the topological charge becomes
\begin{equation}\label{tcexp}
Q = \int_{0}^{\infty} dr q(r),
\end{equation}
where
\begin{equation}\label{qdensity}
\displaystyle q(r) = -\frac{\pi}{4}\cos\left(\frac{\pi}{2}\phi(r)+\delta\right)\frac{d\phi(r)}{dr}
\end{equation}
is the topological charge density. A direct calculation of \eqref{tcexp} reveals that
\be\label{q}
Q=\frac12\sin\Theta(0)-\frac12\sin\Theta(\infty),
\ee
where we have explicitly the relation between the topological charge and the values of (\ref{T}) at its boundaries in the coordinate space.

By construction, the function $\phi (r)$  is dimensionless and, assuming homogeneity throughout the two-dimensional space, we can identify it with a scalar field associated with a Lagrangian in the form
\be\label{model}
{\cal L}=\frac12\dot{\phi}^2-\frac12 \nabla\phi\cdot\nabla\phi-U(\phi),
\ee
where for convenience we write the scalar potential as
\be\label{modelU}
U(r,\phi)=\frac1{2r^2}W^2_{\phi},
\ee
with $W_\phi=dW/d\phi$ for some well-behaved auxiliary function $W=W(\phi)$ \cite{DDR,8}. Thus, exploring potentials of the type (\ref{modelU}) with different properties can bring us new ways to explore the topological properties of the skyrmion-like configurations.

The scalar field we are looking for is the solution of the Euler-Lagrange equations derived from \eqref{model} and \eqref{modelU},  which is given by the second order differential equation
\be\label{modeleq}
r\frac{d}{dr}\left(r\frac{d\phi}{dr}\right)=\frac{1}{2}W_{\phi}W_{\phi\phi}. 
\ee
It is interesting to note that the Lagrangian (5) has energy density $\rho(x)$ which can be written as
\be\label{DE}
\rho(r) = \frac12 \left(\frac{d\phi}{dr}\right)^2+ \frac{1}{2r^2}W^2_{\phi}.
\ee 
In particular, the advantage of writing the potential as it is in (\ref{modelU}) is that we can obtain the minimal energy solutions for the scalar field through the BPS formalism \cite{bps,8}, which is particularly relevant in order to have stable configurations. Such solutions are obtained by solving the first order equation
\be\label{firstorder}
r\frac{d\phi}{dr} = W_{\phi}.
\ee
In this way we have $\rho(r)=\frac{1}{r}\frac{dW}{dr}$ and, as a consequence, the energy we obtain from Lagrangian  (\ref{model})  is  $E=2\pi|\Delta W|$, with $\Delta W = W(\phi(r\rightarrow\infty)) - W(\phi(r=0))$. 

Solutions obtained from (\ref{firstorder}) are stable for radial fluctuations. In fact, scalar field perturbations in the form $\phi(r,t) = \phi(r) + \eta(r)\cos{(w t)}$, with small fluctuation provides a Schrodinger-like stability equation $\hat{H}\eta=w^2 \eta$, which is
\be\label{scheq}
S^{\dag}S\;\eta(r)=w^2\eta,
\ee
with $S= -{d}/{dr} + W_{\phi\phi}/r$ and $S^{\dag}= {d}/{dr} + W_{\phi\phi}/r+1/r$. In this way, the hamiltonian $\hat{H}=S^{\dag}S$, which is a positive operator, has no eigenstates with negative energy, and this ensures stability of the radial solutions against radial fluctuations.  In particular, the groundstate of (\ref{scheq}) can be found by equation $S\eta_0(r)=0$, which can be solved straightforwardly.

In our model the scalar field $\phi $ acts as an auxiliary function so the localized behavior of the energy density is  a consequence of the construction made for the skyrmion system.  Note that the properties of $\phi(r) $, as well as its derivative, are the objects that guide the skyrmion topological properties. Thus, well-behaved solution of the scalar field in a skyrmion configuration contributes to the stability of the magnetic structure. The relevant quantity for our analysis is the topological charge density \eqref{qdensity} since the value of its integral all over space - the topological charge - does not depend on the particular solution of the scalar field but only on the values of $\delta$ and the field at the boundaries $ r = 0 $ and $ r = \infty $.

\subsection{Configurational Entropy} 

Spatially localized structures, in general, have well-behaved energy density in the sense that by being integrated over all space it provides a finite result.  Configurational entropy takes solutions with these characteristics and use them to extract the associated information content. In order to properly define the configurational entropy of some structure, we first define a modal fraction to be worked on. Thus, denoting such quantity by $f({\bf k})$, we have
\be\label{fm}
f({\bf k})= \frac{|F({\bf k})|^2}{\int d^d {\bf k} |F({\bf k})|^2}, 
\ee
where  $F({\bf k})$ is the $ d $-dimensional Fourier transform of the energy density of the solution, denoted by $\rho({\bf x}) $, which is given by
\be\label{FT}
F({\bf k}) = \frac{1}{(2\pi)^{d/2}}\int d^dx \rho({\bf x}) e^{i{\bf k}\cdot{\bf x}}.
\ee
As we are dealing with planar structures, the Fourier transform for $\rho({\bf x})$  must be evaluated for $d=2$. In particular, when the energy density we are studying has only radial dependence, (\ref{FT}) becomes 
\be
F( k) =\int^\infty_0 rdr \rho(r)J_0(kr),
\ee
where  $J_0(kr)$ is the zero-order Bessel Function.

The function $f({\bf k})$ represents a relative weight measured in the $k$-space. Moreover, we make the rescaling $f({\bf k})\to \widetilde{f}({\bf k}) = f({\bf k})/f_{\text{max}}({\bf k})$, where $f_{\text{max}}({\bf k})$ is the maximum modal fraction, to ensure $\widetilde{f}({\bf k})\leq 1$. After these considerations, the configurational entropy (CE) is defined as \cite{gleiser1}
\be\label{sk}
S_C[f]= -\int d^d{\bf k}\,\widetilde{f}({\bf k})\,\text{ln}[\widetilde{f}({\bf k})].
\ee
This quantity has been pointed out as a measure for, say, the {\it quality} of a solution. In general, solutions with higher energies have higher configurational entropy, but even when we have solutions with degenerate energies, $S_C[f]$ can act as a tie-breaking factor in choosing the most appropriate one. Here we call 
\begin{equation}
{\cal S}({\bf k})=-\widetilde{f}({\bf k})\text{ln}[\widetilde{f}({\bf k})]
\end{equation}
the configurational entropy density, which is used to describe how the configurational entropy of a given structure is distributed in the functional space.

A key point in describing a skyrmion is the relationship between the existence of its topological charge and its stability. In other words, the skyrmion is a magnetic structure with topological protection due to the existence of the charge (\ref{tc}), which makes it stable. The topological charge is independent of the variation in the parameters of the models we analyze in this work, since we always have unit or half-unit charge, so the only physical entity that supports informational content is $q(r)$. Thus, when looking for different ways of measuring qualitatively how information can be evaluated in a given set of solutions, it is more interesting that instead of looking at the energy density of the scalar field solution, we look at the topological charge density (\ref{qdensity}). In this way, we are slightly changing the perspective about the original approach given in \cite{gleiser1}, since we are now relating topological and informational features  of the structure under consideration. We perform this procedure by changing
\begin{equation}
\rho(r)\to q(r)
\end{equation}
in Eq.~ \eqref{FT}. Here we note that this modification does not represent any aggression to the original proposal since what is necessary for this type of analysis is the localized character of the structure being studied, where one must look for quantities that are square integrable so that we can Fourier transform it. In this sense the appropriate behavior of the topological charge density is guaranteed by the localized behavior of the energy density of the scalar field.

Configurational entropy is an approach derived from information theory. In this way one can infer that the structure of the skyrmion-like model we study here can be seen as being composed by a transmitter (system), a channel (equations of motion + boundary conditions) and a receiver (spatially localized structure). Schematically we have:
$$
\xymatrix{ *+[F]{\underbracket[0pt][0pt]{\text{System}}_{\text{(transmitter)}}}\ar[r]& *+[F-]{\underbracket[0pt][0pt]{\text{EoM+BC}}_{\text{(receiver)}}}\ar[r]& *+[F-]{\underbracket[0pt][0pt]{\text{Spat. Loc. Structure}}_{\text{(receptor)}}}
}
$$
From this point of view one can interpret $S_C[f]$ as the measure of the informational content of spatially localized solutions by analogy to Shannon entropy. Shannon entropy does not provide specific information about which message the transmitter sent since it only depends on the probability distribution generated by the set of possible messages to be sent. In the same way, the configurational entropy does not reveal information about which specific solution it provided, but it only addresses the profile of the topological charge distribution of the skyrmion through its map on the functional space.

\begin{figure*}[t]
\center
\subfigure[\,The skyrmion. \label{fig1a}]{\includegraphics[scale=0.3]{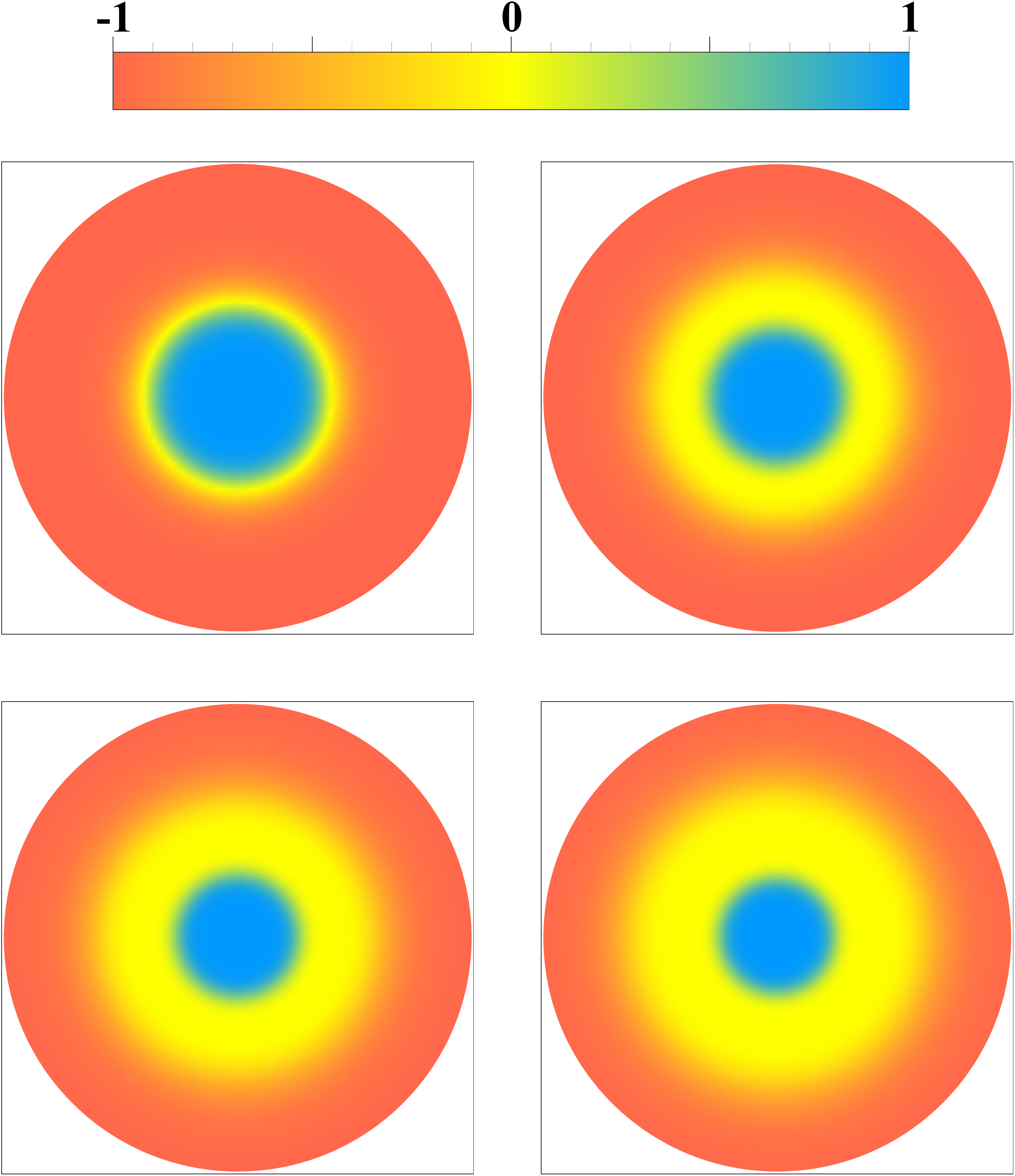}}
\hspace{1cm}
\subfigure[\,The topological charge density. \label{fig1b}]{\includegraphics[scale=0.3]{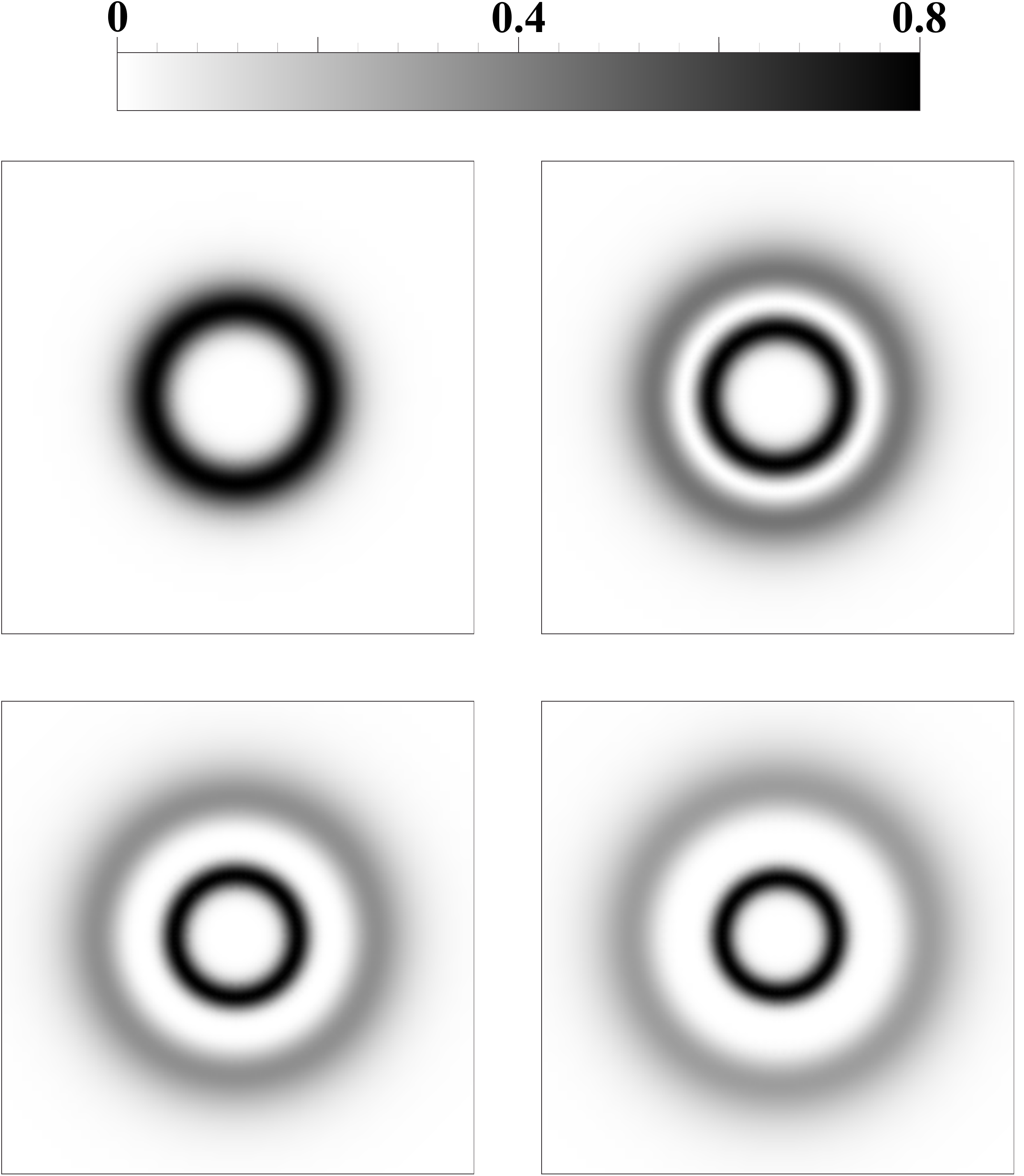}}
\caption{(a) The topological structure with skyrmion number $Q=1$ and (b) the topological charge density \eqref{q1} for the model \eqref{p1}, depicted for $s=0.6$ and $p=1,3,5,7$, from top left to bottom right in both the (a) and (b) figures.}
\label{fig1}
\end{figure*}

\section{Models} 

Here, we aim to introduce the concept of configurational entropy in the context of skyrmions. For this reason, we first review a model introduced in \cite{jmmm}, which has analytical solutions, and then investigate a new model, which generalizes a model studied in \cite{DDR}. In the two cases, we bring novelties related to the subject of this work, which concerns the conformational entropy of the topological structures they support.  

\subsection{Skyrmion} 

The first model we start studying was presented in \cite{jmmm} and has as $W$-function the polynomial given by 
\begin{equation}
W_p(\phi) = \frac{p^2\,\phi^{(2p-1)/p}}{(1-s)(2p-1)} - \frac{p^2\,\phi^{(2p+1)/p}}{(1-s)(2p+1)}
\end{equation}
where $s\in[0,1)$ and $p=1,3,5,7,..., ~$ is an odd integer. The  $W-$function obeys $W_\phi = \frac{p}{(1-s)}\phi(\phi^{-1/p}-\phi^{1/p})$, and provides the following scalar potential
\be\label{p1}
U(\phi;r)= \frac{1}{2r^2}\frac{p^2}{(1-s)^2}\phi^2(\phi^{-1/p}-\phi^{1/p})^2.
\ee
Note that for $ p = 1 $  we have the $\phi^4$ model and there are two degenerate minima at $\bar{\phi}=\pm 1$. The $s$ parameter is introduced in the system to control the energy of the system. The scalar field solution is obtained by solving (\ref{firstorder}), in the form
\be \label{phip}
\phi_{p,s}(r)=\left(\frac{1-r^{2/(1-s)}}{1+r^{2/(1-s)}}\right)^p.
\ee
For all values of $p$ and $s$ we have $(\phi_{p,s}(0),\phi_{p,s}(\infty))=(1,-1)$. Plugging solution \eqref{phip} into magnetization (\ref{M}) one finds a topological skyrmion where its core is filled with the magnetization pointed upward and as one moves away from that region the direction of the magnetization varies until it get to the border of the defect, where it points downward. The shape of the skyrmion for some fixed values of $p$ and  $s$  is depicted in Fig.~\ref{fig1} (a), where we can observe that the higher $p$, the higher the yellow area covered by $\phi\approx0$, indicating a kind of internal structure for the skyrmion.

\begin{figure}[t]
\centerline{\includegraphics[height=15.8em]{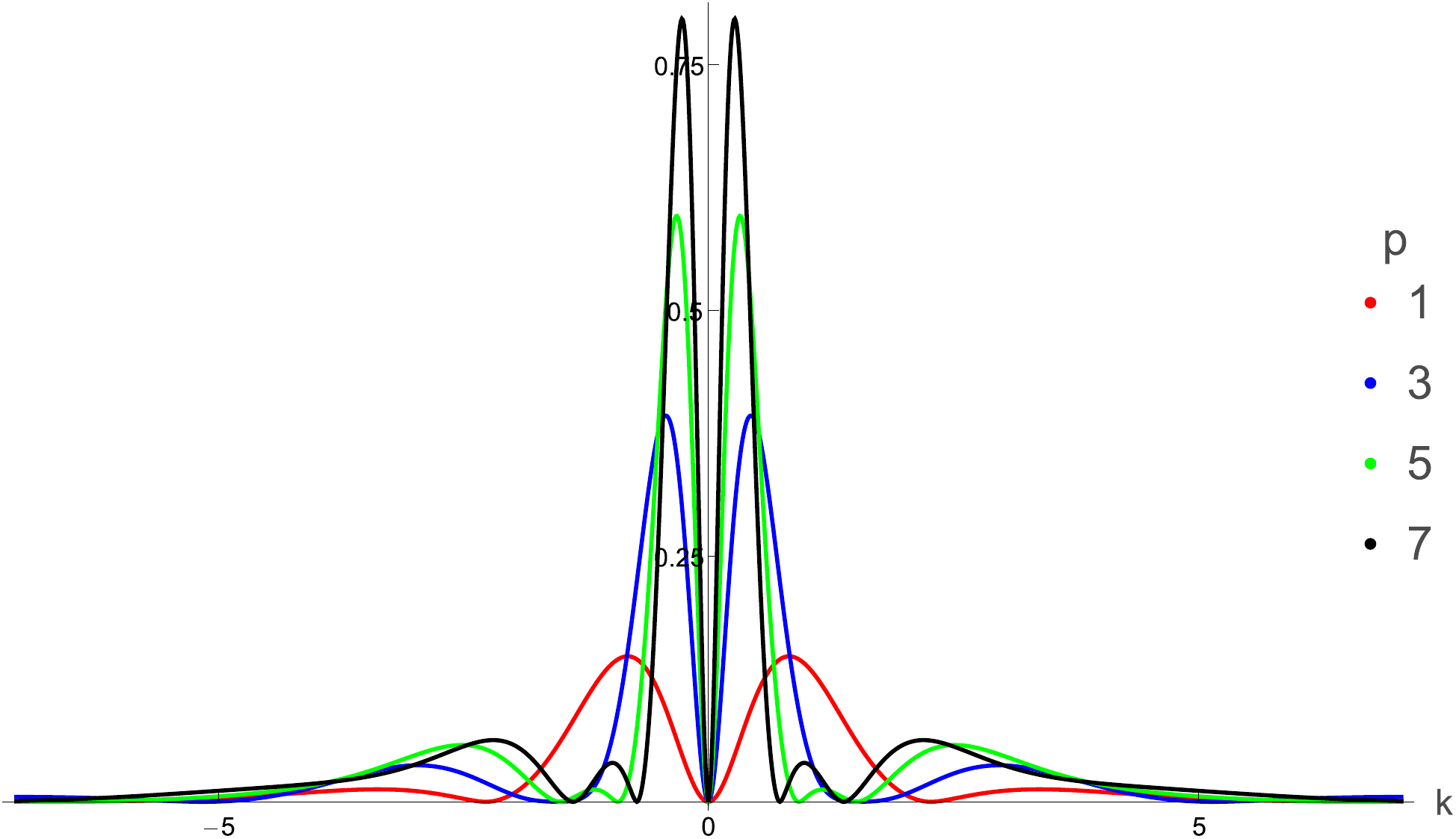}}
\caption{Configurational entropy density ${\cal S}$ for the skyrmion model \eqref{p1} for some values of $p$ and $s = 0$.}\label{fig2}
\end{figure}
\begin{figure}[t]
\centerline{\includegraphics[height=15.8em]{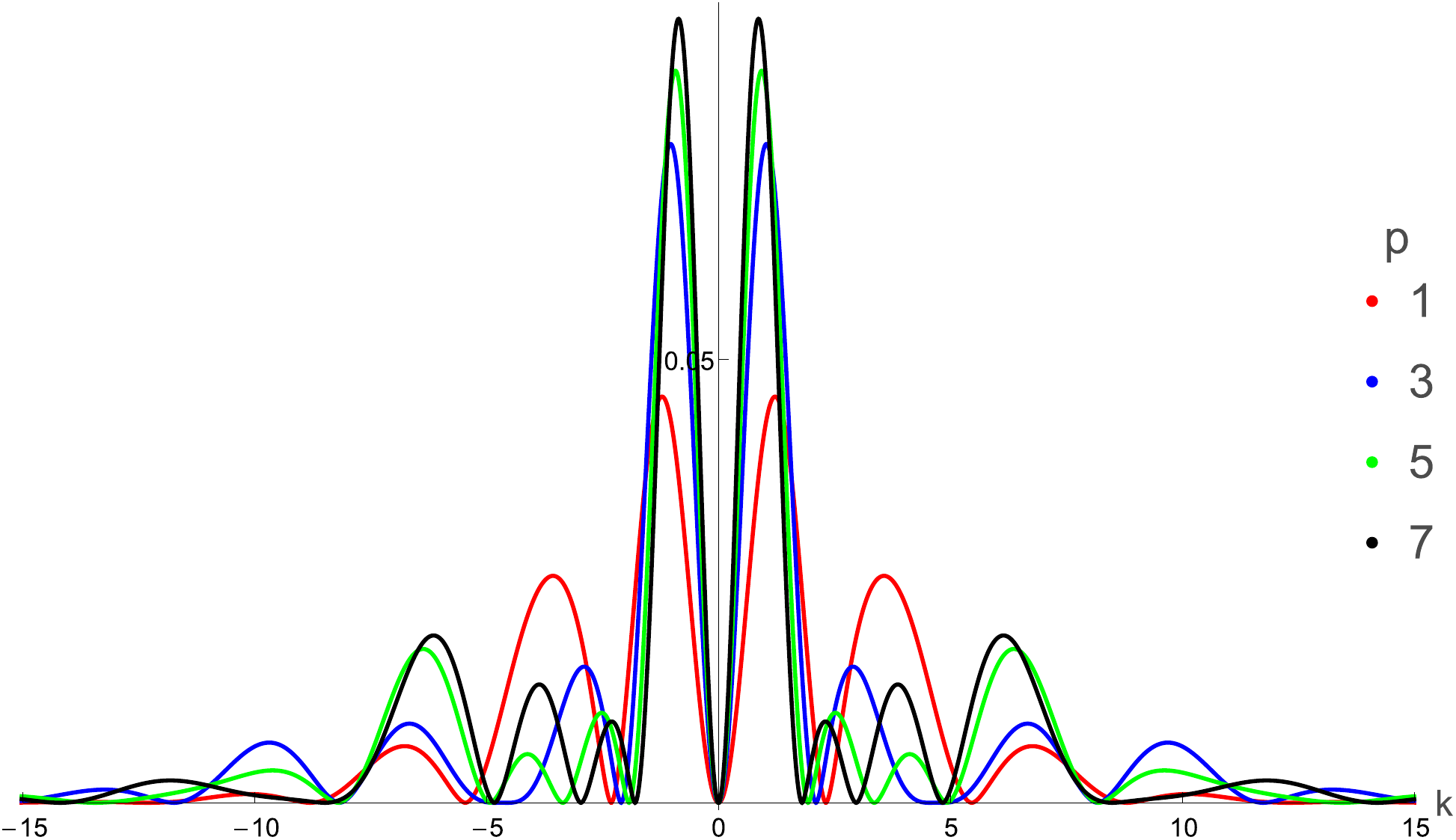}}
\caption{Configurational entropy density ${\cal S}$ for the skyrmion model \eqref{p1} for some values of $p$ and $s=0.6$.}\label{fig3}
\end{figure}
\begin{figure}[t]
\centerline{\includegraphics[height=22em]{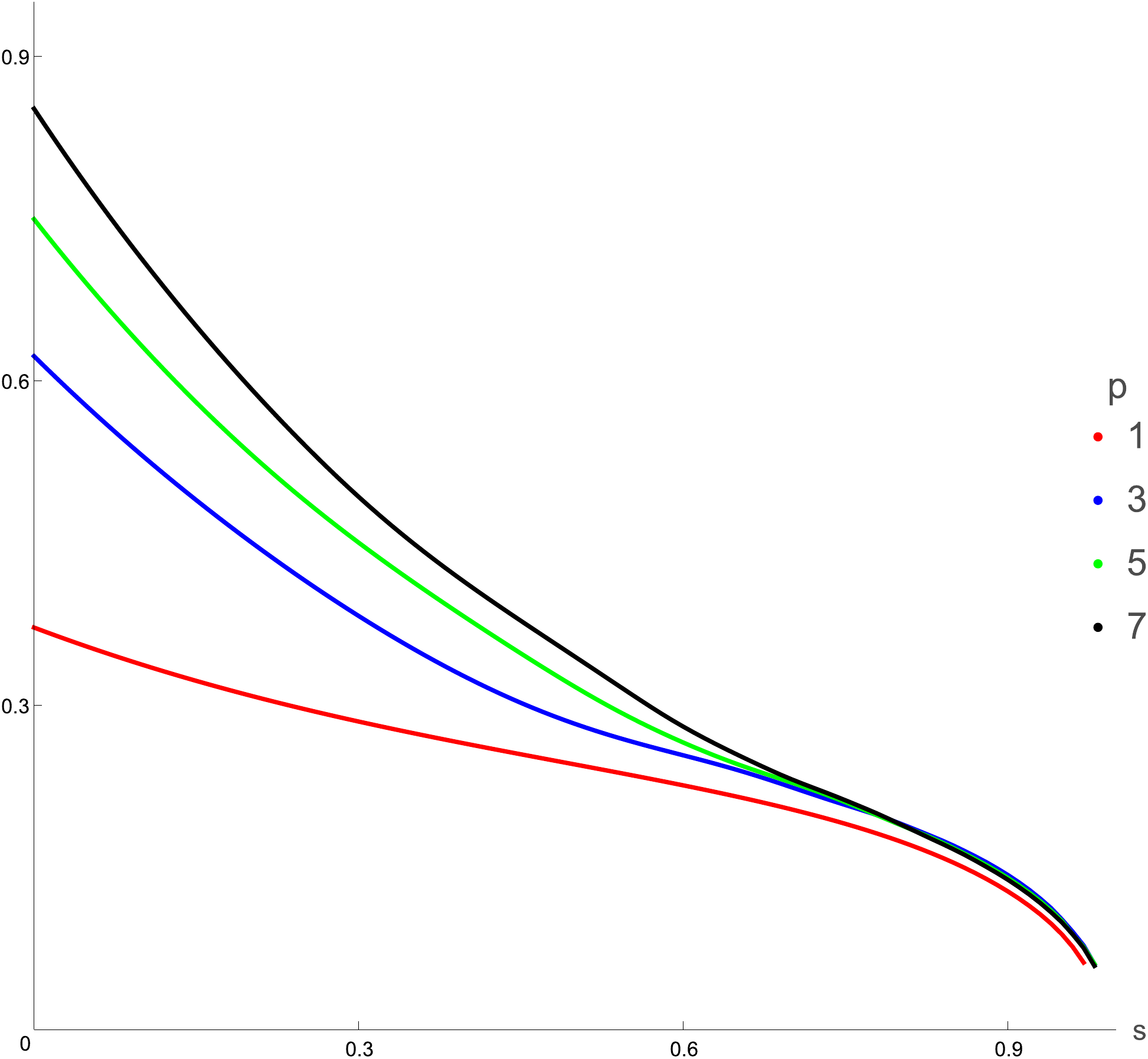}}
\caption{The configurational entropy $S_c$ of the model \eqref{p1} for some values of $p$.}\label{fig4}
\end{figure}

Since we have the expression for the scalar field of the model, we directly obtain the topological charge density \eqref{q}, which for solution \eqref{phip} is given by
\be\label{q1}
q_{p,s}(r)= q_0(r)\cos\frac{\pi}{2}\left(\frac{1-r^{2/(1-s)}}{1+r^{2/(1-s)}}\right)^p,
\ee
where
\be
q_0(r)=\frac{p\pi}{(1-s)}  \frac{\left(1-r^{2/(1-s)}\right)^{p-1}}{\left(1+r^{2/(1-s)}\right)^{p+1}} r^{(1+s)/(1-s)}, 
\ee
and its shape is depicted in Fig.~\ref{fig1} (b) for some values of $p$ and fixed  $s=0.6$. We can observe that $ p $ acts on the defect with the formation of {\it halos} in the topological charge density, which disperses on the plane as $ p $ increases. This is due to the fact that the scalar field has an internal structure characterized by the presence of a point, different from the boundaries, where the derivative of the field is zero. So the presence of these halos are consequences of the splitting-like effect in the topological charge density. The parameter $p$ controls both position and amplitude of the splitting,  since it is related with the order of the derivative of the field for which we have a nonzero value at the point where is situated the internal structure. The $s$ parameter acts to make the transition between the upward and downward states of the magnetization smoother and, as a consequence, it also acts on the intensity of the halos in the topological charge density.

Unfortunately, it was not possible to find analytical results for the modal fraction or configurational entropy density associated to the system topology, so we performed a numerical analysis of these quantities.  In particular, we work directly on the density of configurational entropy, which is depicted in Figs.~\ref{fig2} and \ref{fig3}. In Fig.~\ref{fig2}  we have the distribution of the configurational entropy density in $k$-space for some values of $p$ and $s=0$. We can observe a symmetric distribution around $k=0$ with the presence of a series of hills that, as $p$ increases, becomes higher and more localized. Note that the number of hills is the same for any value of $p$, so the entropy distribution only responds to the increase of the exponent of the scalar field in a quantitative way, without qualitative changes in the shape of the distribution. The quantitative character of the configurational entropy is shown in Table I for some values of $s$ and $p$.

\begin{table}[h!]
\centering
\small
\caption[]{Values of the configurational entropy for the model \eqref{p1} for several values of $s$ and $p$.}
\begin{tabular}{ccccc}
\hline
\hline
\begin{minipage}[t]{.07\textwidth}\begin{flushleft} $ $\end{flushleft}\end{minipage}&
\begin{minipage}[t]{.09\textwidth} $p=1$\end{minipage}&
\begin{minipage}[t]{.09\textwidth} $p=3$\end{minipage}&
\begin{minipage}[t]{.09\textwidth} $p=5$\end{minipage}&
\begin{minipage}[t]{.09\textwidth} $p=7$\end{minipage}\\
\hline
$s=0.0$ &$0.37$	&$0.62$ &$0.75$ &$0.85$ \\
\hline
$s=0.2$ &$0.31$	&$0.45$ &$0.53$ &$0.59$ \\
\hline
$s=0.4$ &$0.26$	&$0.33$ &$0.38$ &$0.41$ \\
\hline
$s=0.6$ &$0.23$	&$0.25$ &$0.27$ &$0.28$ \\
\hline
$s=0.8$ &$0.17$	&$0.19$ &$0.19$ &$0.19$ \\
\hline
\hline
\end{tabular}
\label{tab1}
\end{table}

In Fig.~\ref{fig3} we display ${\cal S}({\bf k})$ for some values of $p$ and $s=0.6$.  One can observe an increase in the number of hills in this case and, additionally, that these hills are smaller than those shown in Fig.~\ref{fig2}. It occurs because the  modal fraction responds to the increasing of $s$ with an increasing in the number of points where $f({\bf k})=0$ and a gradual diminution of its height, together with a scattering of the structure in the Fourier space. In Fig.~\ref{fig4} we show the behavior of $S_c$ for some values of $p$. We notice that as $s$ approaches unity the informational content of the system becomes less expressive, decreasing gradually when $s$ increases, as also presented in Table I.

\begin{figure*}[t]
\center
\subfigure[\,The vortex. \label{fig5a}]{\includegraphics[scale=0.3]{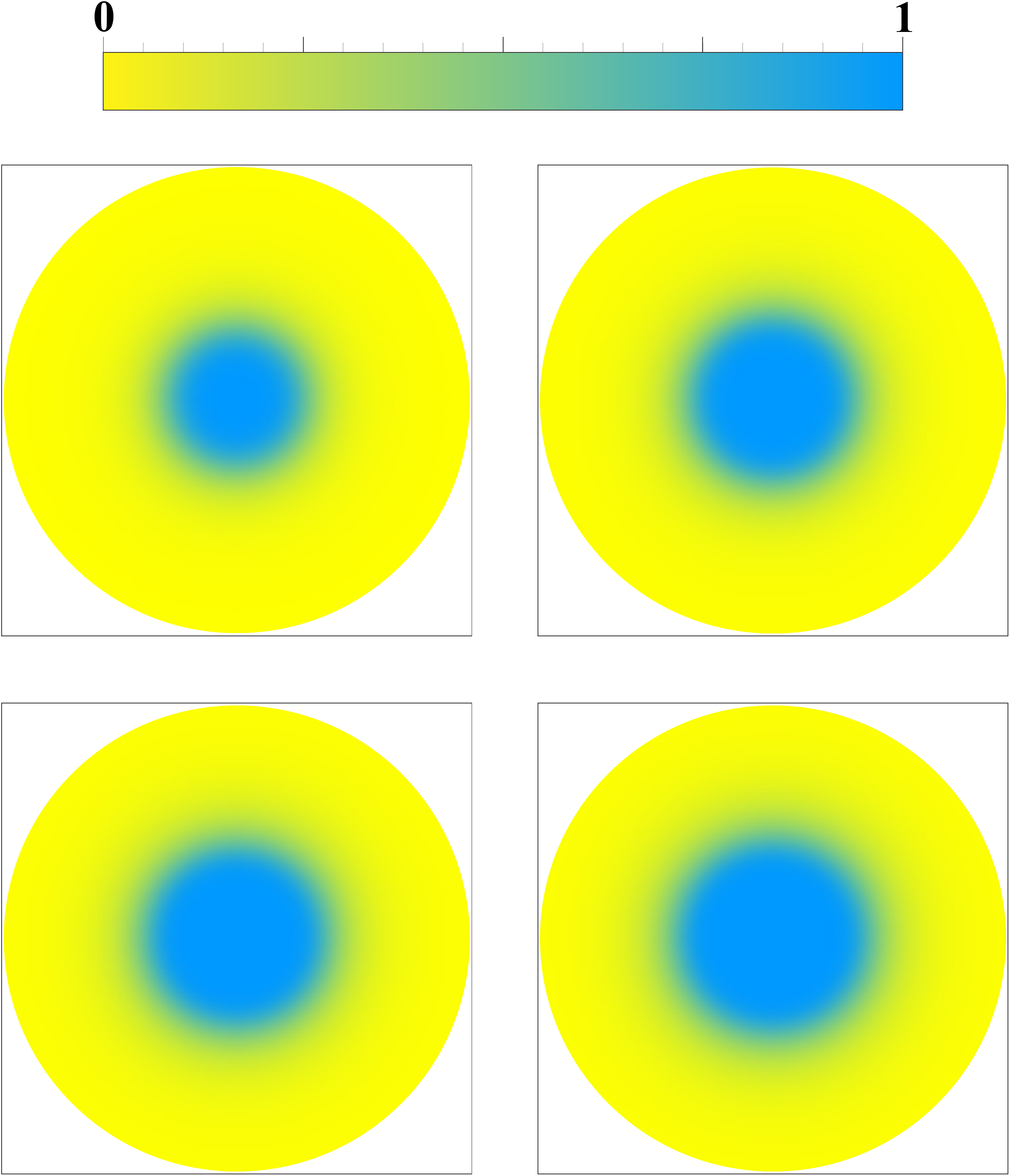}}
\hspace{1cm}
\subfigure[\,The topological charge density. \label{fig5b}]{\includegraphics[scale=0.3]{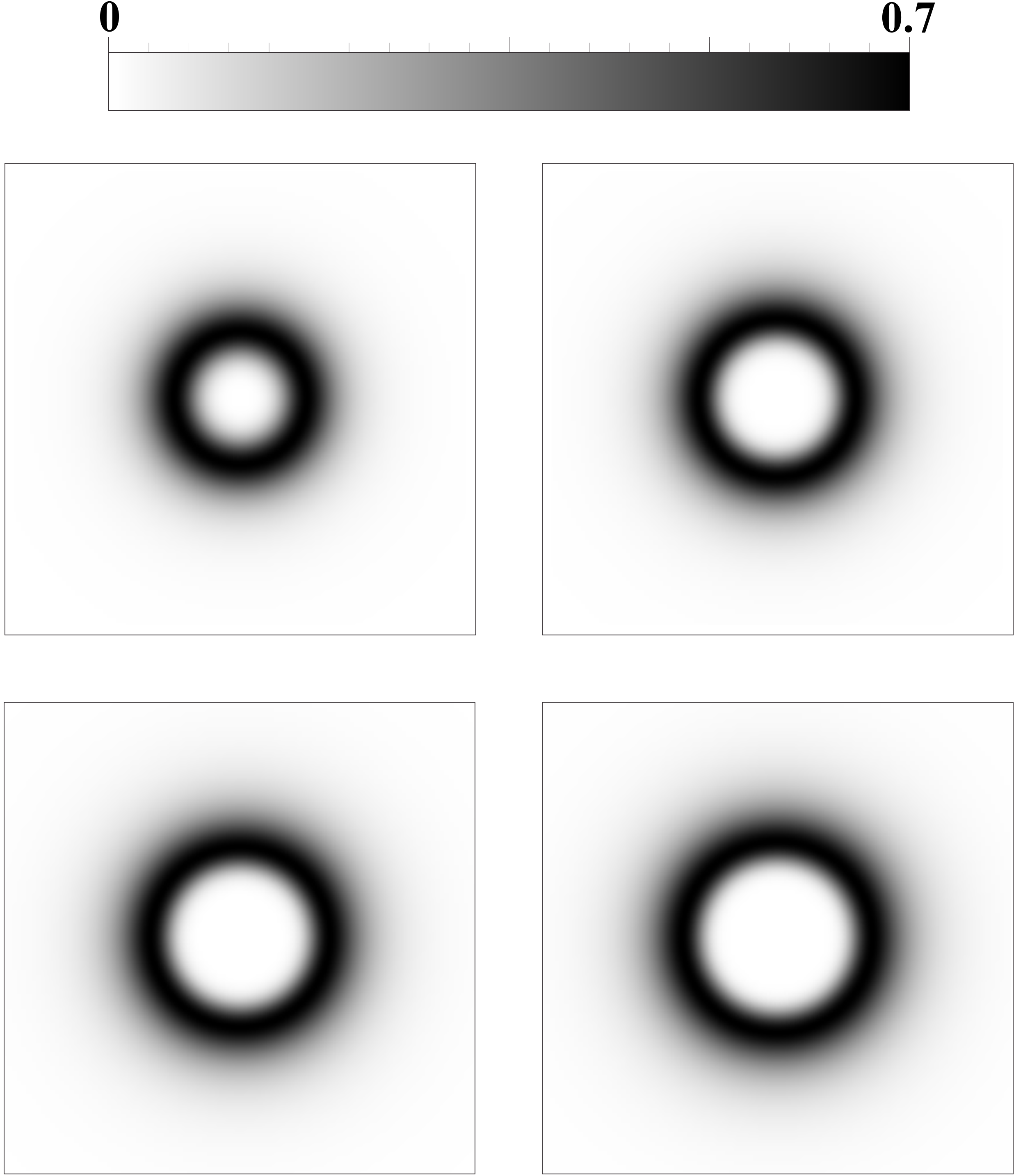}}
\caption{(a) The topological structure with skyrmion number $Q=1/2$ which is controlled by the model \eqref{p2} and (b) the corresponding topological charge density \eqref{q2}, depicted for $s=0.6$ and $n=1,2,3,4$, from top left to bottom right in both (a) and (b) figures.}
\label{fig5}
\end{figure*}

\subsection{Vortex}

We now continue our study  presentating a new model that has a vortex-like behavior. As discussed in the introduction, such solutions are characterized by the existence of a fractional topological charge. The model we are interested in is generated by the polynomial function 
\begin{equation}
W_n(\phi) = \frac{n\,\phi^2}{2(1-s)} -\frac{n^2\,\phi^{2(n+1)/n}}{2(n+1)(1-s)},
\end{equation}
where we have $s\in [0,1)$ and $n=1,2,3,4,...,$ is a positive integer. The derivative of this function is $W_{\phi}=\frac{n}{(1-s)}\phi(1-\phi^{\frac{2}{n}})$, so the scalar potential becomes
\be\label{p2}
U(\phi,r) = \frac{1}{2r^2}\frac{n^2}{(1-s)^2}\phi^2(1-\phi^{\frac{2}{n}})^2. 
\ee
This potential has degenerate minima at $\bar{\phi}_{\pm} = \pm 1$ and $\bar{\phi}_0 = 0$ for any value of $n$. Note that for $n=1$  we retrieve the model studied in \cite{DDR}. In particular, for our study we use the solution connecting $\bar{\phi}_0 = 0$ and $\bar{\phi}_{+} = 1$, which is given by
\be\label{sol2}
\phi_{n,s}(r)=\frac{r^{n/(1-s)}}{(1+r^{2/(1-s)})^{n/2}}.
\ee
We have $\phi_{n,s}(0)=0$ and $\phi_{n,s}(\infty)=1$, so one must be careful with the choice for $\delta$ in the magnetization vector (\ref{M}). Plugging $\delta=\pi/2$ and solution (\ref{sol2}) in formula (\ref{M})  we obtain a magnetization vector that points in the
$\hat{z}$-direction at the origin and in the $\hat{\theta}$-direction at infinity. As a consequence, the topological charge is given by $Q= 1/2$, which characterizes a vortex-like solution. The effect of increasing the $ n$ parameter in the model is presented in Fig.~\ref{fig5} (a) for a fixed value of
$s$, where we can see that as $n$ gets bigger, the core of the skyrmion, where the magnetization points upward, increases; see also Fig. \ref{fig5}(b) for the behavior of the topological charge density, which is now given by
\be\label{q2}
q_{n,s}(r)=q_0(r)\sin\frac{\pi}{2}\left(\frac{r^{n/(1-s)}}{(1+r^{2/(1-s)})^{n/2}}\right),
\ee
with
\be
q_0(r)=\frac{n\pi}{4(1-s)}\frac{r^{(n+s-1)/(1-s)}}{\left(1+r^{2/(1-s)}\right)^{(n+2)/2}}.
\ee
It responds to this variation with an increasing of the core sector, where $q(r) = 0$. Note that now we do not have the presence of halos like those that appeared in the skyrmion model studied previously. This is due to the fact that now the solution of the scalar field has no internal structure, so there is no splitting-like behavior in the topological charge density.
\begin{figure}[t]
\centerline{\includegraphics[height=15.8em]{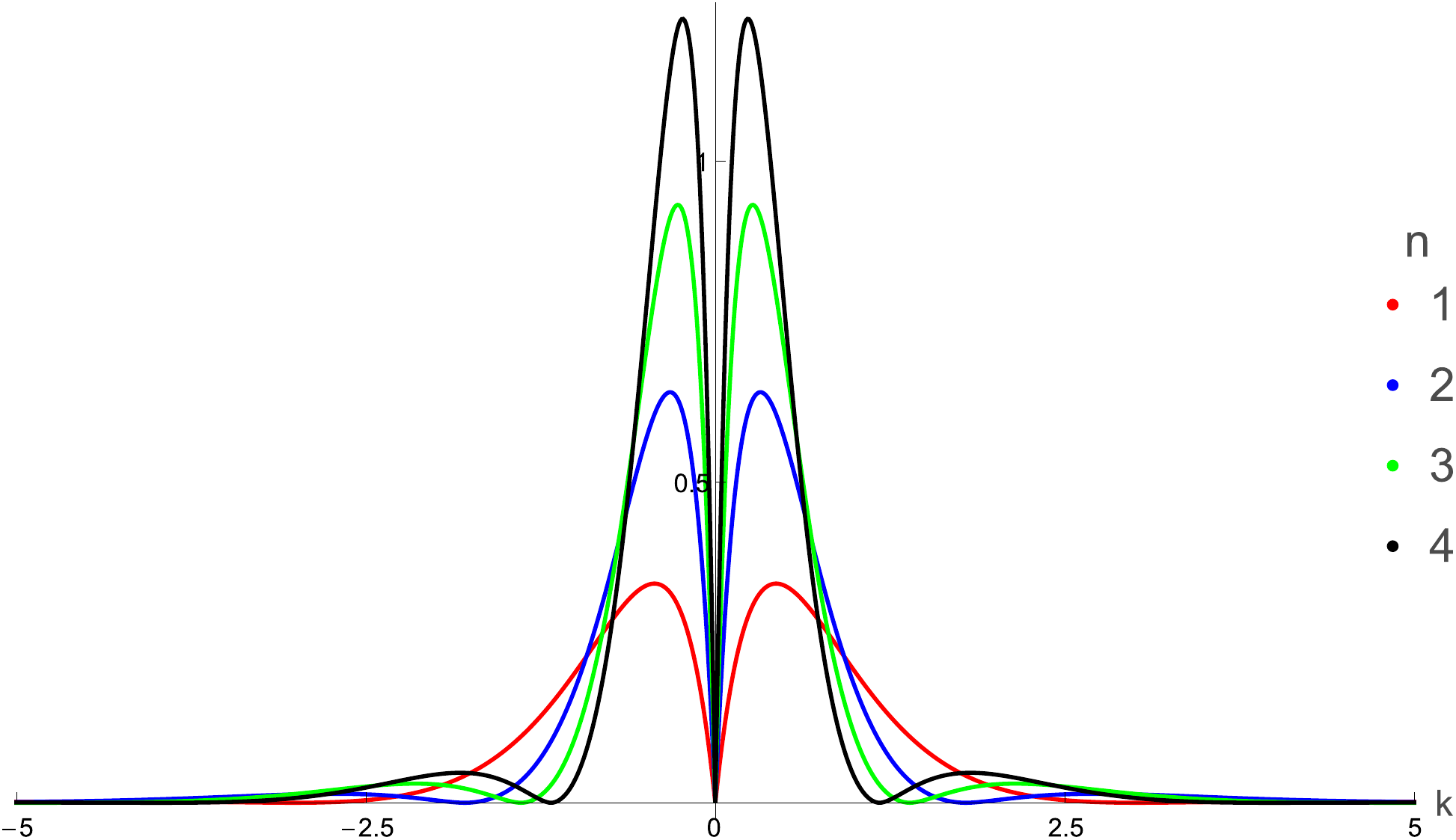}}
\caption{Configurational entropy density ${\cal S}$ for the skyrmion model \eqref{p2} for some values of $ n $ and $ s = 0. $}\label{fig6}
\end{figure}
\begin{figure}[t]
\centerline{\includegraphics[height=18em]{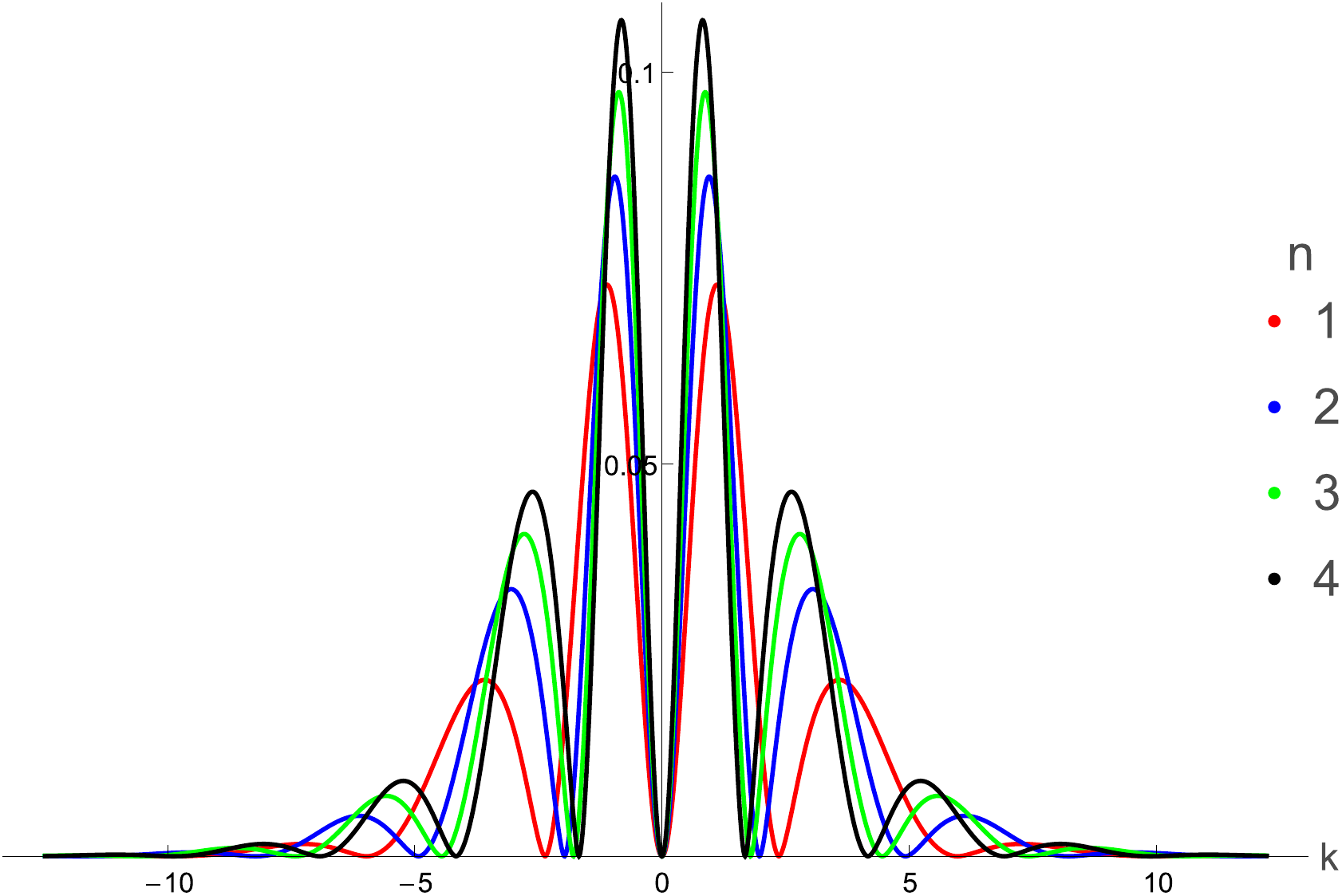}}
\caption{Configurational entropy density ${\cal S}$ for the skyrmion model \eqref{p2} for some values of $ n $ and $ s = 0.6. $}\label{fig7}
\end{figure}

The $ s $ parameter again acts as a control factor for the energy and the smoothness of the transition between the magnetization in the $\hat{z}$ and $\hat{\theta}$  directions. However, it is natural to expect that it also has relevance when characterizing the size of the defect. In order to observe this relationship we use the energy density of the scalar field, given by
\be\label{ed2}
\rho_{n,s}(r)= \frac{n^2}{(1-s)^2}\frac{ r^{2(n+s-1)/(1-s)}}{(1+r^{2/(1-s)})^{n+2}}, 
\ee
to define the quantity $R_{n,s} = \bar{r}_{n,s}/\bar{r}_{1,0}$, where we use $\bar{r}_{n,s} = \left(\int_0^{\infty}\rho_{n,s}(r)r^2dr\right)/\left(\int_0^{\infty}\rho_{n,s}(r)rdr\right)$. This quantity was introduced in \cite{DDR} as a proposal to measure the skyrmion size. In particular, for the model we are studying, one has
\be
R_{n,s} = \frac{n(n+1)\Gamma((2n-s+1)/2)}{\Gamma(n+2)\Gamma((3-s)/2)},
\ee
where $\Gamma(\xi)$ is the Gamma Function. In fact, the size of the defect is related to both $ n $ and $ s $. One can note that $ R_{n,s} $ increases as $ n$ increases and decreases as $ s $ increases, except for $n=1$, where we have $R_{1,s}=1$.

\begin{figure}[t!]
\vspace{0.3cm}
\centerline{\includegraphics[height=20em]{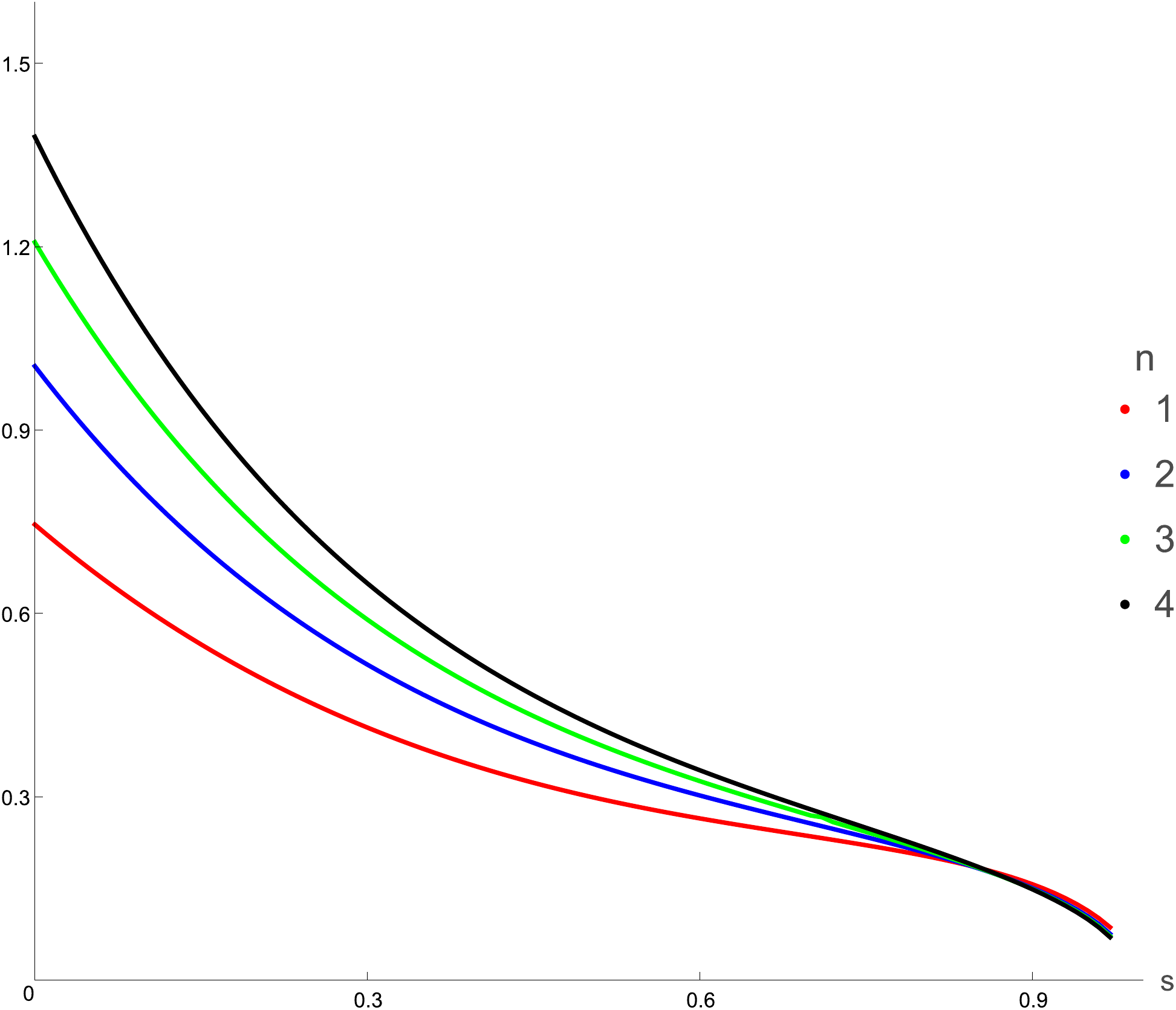}}
\caption{The configurational entropy $S_c$ of the model \eqref{p2} for some values of $ n $.}\label{fig8}
\end{figure}

After presenting some basic features of the model, let us now explore its informational behavior. Unfortunately, again we could not obtain analytical solutions, so we performed a numerical analysis of the problem. By inserting \eqref{q2} into \eqref{fm} we obtain the modal fraction associated with this system and, with this, we calculate the configurational entropy density that is shown for some values of $n$ and $s=0$ in Fig.~\ref{fig6}. Note that we have hills that grow and become more localized as the value of $n$ increases. Also, in Fig.~\ref{fig7} one shows the configurational entropy density for the same model, but with $s=0.6$, where we can observe the increase in the number of hills and a significant decrease in its height. Thus, there is some similarity with the skyrmion model analyzed in the previous section due to the polynomial format of the potentials \eqref{p1} and \eqref{p2} and, consequently, the discussion about the parameters $s$ and $n$ of the model \eqref{p2} follows in the same direction of what was presented for the parameters $s$ and $p$ in \eqref{p1}. In Table II we display values of the configurational entropy for some values of $s$ and $n$ and in Fig.~\ref{fig8} we show the behavior of $S_c$ for some constant values of $n$.  We observe that the increasing of $n$ contributes to the increasing of the configurational entropy, while $s$ contributes to its decreasing.

\begin{table}[t!]
\centering
\small
\caption[]{The configurational entropy for the model \eqref{p2} for some values of $n$ and $s$.}
\begin{tabular}{ccccc}
\hline
\hline
\begin{minipage}[t]{.07\textwidth}\begin{flushleft} $ $\end{flushleft}\end{minipage}&
\begin{minipage}[t]{.09\textwidth}$n=1$\end{minipage}&
\begin{minipage}[t]{.09\textwidth}$n=2$\end{minipage}&
\begin{minipage}[t]{.09\textwidth}$n=3$\end{minipage}&
\begin{minipage}[t]{.09\textwidth}$n=4$\end{minipage}\\
\hline
$s=0.0$ &$0.75$	&$1.00$ &$1.21$ &$1.38$ \\
\hline
$s=0.2$ &$0.50$	&$0.64$ &$0.74$ &$0.82$ \\
\hline
$s=0.4$ &$0.35$	&$0.42$ &$0.48$ &$0.52$ \\
\hline
$s=0.6$ &$0.26$	&$0.30$ &$0.33$ &$0.34$ \\
\hline
$s=0.8$ &$0.20$	&$0.21$ &$0.21$ &$0.22$ \\
\hline
\hline
\end{tabular}
\label{tab3}
\end{table}


\section{Ending comments}
In this work we investigated the configurational entropy associated to planar topological structures of the skyrmion type. We analyzed the configurational entropy associated to the topological number density which is in general used to describe planar skyrmions in magnetic materials at the nanometric scale. 

After reviewing the basic facts on the models used to describe skyrmions and the configurational entropy, we worked out two distinct models, the first one supporting skyrmion-like configurations with unity skyrmion number, controlled by the parameters $s$ and $p$, with $s\in[0,1)$ and $p$ a non-negative odd integer. The second model is different, and supports half-skyrmion or vortex-like configurations with skyrmion number 1/2, controlled by the parameters $s$ and $n$, with $n$ being a non-negative integer.\\

The results showed that, even though dealing with solutions with the same topological charge, the different forms of the topological charge density lead to distinct features for the configurational entropy density and to distinct values for the configurational entropy. In particular, the increasing of the parameter $s$ contributes to diminish the configurational entropy, although the other parameters $p$ and $n$ act differently, increasing the configurational entropy as they increase.

\acknowledgments{The authors would like to thank CNPq for partial financial support.}


\end{document}